\documentclass[aps,prb,twocolumn,10pt,superscriptaddress,amsmath,amssymb]{revtex4-1}

\usepackage{graphicx}
\usepackage{subfigure}
\usepackage{epstopdf}
\usepackage{hyperref}
\usepackage{slashed}
\usepackage{bm}
\usepackage{feynmp}

\def\nablab{{\mbox{\boldmath $\nabla$}}}

\begin{document}

\title{Non-local topological magnetoelectric effect by Coulomb interaction at a topological insulator-ferromagnet interface}

\author{Stefan Rex}
\affiliation{Department of Physics, Norwegian University of Science and Technology, N-7491 Trondheim, Norway}
\author{Flavio S. Nogueira}
\affiliation{Institute for Theoretical Solid State Physics, IFW Dresden, PF 270116, 01171 Dresden, Germany}
\affiliation{Institut f{\"u}r Theoretische Physik III, Ruhr-Universit\"at Bochum,
	Universit\"atsstra\ss e 150, DE-44801 Bochum, Germany}
\author{Asle Sudb{\o}}
\affiliation{Department of Physics, Norwegian University of Science and Technology, N-7491 Trondheim, Norway}

\date{\today}

\begin{abstract}

The interface between a topological insulator and a ferromagnetic insulator exhibits an interesting interplay of topological Dirac electrons and magnetism. As has been shown recently, the breaking of time-reversal invariance by magnetic order generates a Chern-Simons term in the action, that in turn leads to a Berry phase and a magnetoelectric effect of topological origin. Here, we consider the system in the presence of long-range Coulomb interaction between the Dirac electrons, and find that the magnetoelectric effect of the fluctuating electric field becomes non-local. We derive a Landau-Lifshitz equation for the fluctuation-induced magnetization dynamics and the Euler-Lagrange equation of the Coulomb field by explicit one-loop calculations. Via the Coulomb interaction, divergences in the in-plane magnetization affect the magnetization dynamics over large distances in a topologically protected way.
\end{abstract}

\maketitle

\section{Introduction}\label{SecIntroduction}
In a topological insulator (TI), the bulk band structure gives rise to gapless surface states that are protected by symmetry via a bulk-boundary correspondence.\cite{Kane-RMP,Zhang-RMP}  
These conducting states have a linear dispersion (Dirac electrons) 
 arising mainly due to strong spin-orbit coupling. In addition, spin-momentum locking makes surface currents on a 
 TI a promising tool for spintronics applications.\cite{QiHughesZhang_Nature, YokoyamaTanakaNagaosa} 
 However, not all materials that feature a Dirac dispersion and a strong spin-orbit coupling 
 are TIs. For instance, pure bismuth is a Dirac-like material featuring a strong spin-orbit coupling, which is not a TI, since its  
 surface states are not protected by symmetry. The protecting symmetry in most TIs is time-reversal invariance (TRI). 

In three-dimensional (3D) TIs, the electromagnetic response is characterized by a magnetoelectric term in the Lagrangian.\cite{QiHughesZhang, EssinMV} 
Unlike the magnetoelectric term arising in other materials, like for example, multiferroics, the magnetoelectric term in 
TI electrodynamics is intrinsically topological, both due to the topological properties in reciprocal lattice space, and in real space. 
This can be seen by applying an external magnetic field perpendicular to the surface of a 3D TI of thickness $L$. A  computation of the vacuum polarization of two-dimensional Dirac fermions in 
the presence of an external field for each TI surface, yields the action \cite{Semenoff,Redlich}
\begin{eqnarray}
\label{Eq:diff-CS}
S_{\rm vpol}&=&\frac{e^2}{8\pi}\int\!\text{d}t\int\!\text{d}x\,\text{d}y\,\varepsilon_{\mu\nu\lambda}\nonumber\\
&&\quad\quad\left(A^\mu\partial^\nu A^\lambda|_{z=L}
- A^\mu\partial^\nu A^\lambda|_{z=0} \right),
\end{eqnarray}
where we have adopted a covariant notation and $z=0$ and $z=L$ correspond to the lower and 
upper surfaces, respectively. We work in units where $c=1$ and $\hbar=1$. The above action 
yields the difference between Chern-Simons (CS) terms generated by the vacu\-um polarization 
on both surfaces. It can rewritten as the integral of a total derivative,
\begin{eqnarray}
\label{Eq:top-term}
S_{\rm vpol}&=&\frac{e^2}{16\pi}\int\!\text{d}t\int\!\text{d}x\,\text{d}y\int_{0}^{L}\!\!\text{d}z\,\,
\partial_z\left(\varepsilon_{\mu\nu\lambda}A^\mu F^{\lambda\nu}\right)
\nonumber\\
&=& \frac{e^2}{32\pi}\int\!\text{d}^4x\,\varepsilon_{\mu\nu\lambda\rho}F^{\mu\nu} F^{\lambda \rho}, 
\end{eqnarray}
where, in passing from the first to the second line, the expression has been made fully covariant. The above equations
follow from the assumption that the Fermi level of each surface state lies precisely at zero, i.e., at the Dirac point of the Dirac spectrum. Moreover, spin-momentum locking implies that the Dirac fermions at the upper surface have a helicity opposite to the lower ones. Thus, we have obtained a magnetoelectric term that is overall TRI.  A more general form is given by 
\begin{equation}
\label{Eq:theta-term}
S_{\rm vpol}=\frac{e^2\theta}{32\pi^2}\int\!\text{d}^4x\,\varepsilon^{\mu\nu\sigma\tau} F_{\mu\nu}F_{\sigma\tau},
\end{equation} 
where $\theta$ is given by\cite{QiHughesZhang}
\begin{equation}
\label{Eq:theta}
\theta=\frac{1}{8\pi}\int\!\text{d}^3k~{\rm tr}\left[{\bf a}({\bf k})\wedge {\bf f}({\bf k})-\frac{2}{3}{\bf a}({\bf k})\wedge {\bf a}({\bf k})\wedge {\bf a}({\bf k})\right]
\end{equation}
where the 2-form ${\bf f}({\bf k})$ yields the Berry curvature,
\begin{equation}
{\bf f}({\bf k})=\mathrm{d}{\bf a}({\bf k})+i{\bf a}({\bf k})\wedge {\bf a}({\bf k}),
\end{equation}
with
\begin{equation}
{\bf a}_{\alpha\beta}({\bf k})=-i\langle\alpha,{\bf k}|\nablab_{\bf k}|\beta,{\bf k}\rangle,
\end{equation}
being the non-abelian Berry vector potential associated with the Bloch 
state $|\alpha,{\bf k}\rangle$.  Thus, the electromagnetic response 
of 3D TIs yields an interesting interplay between the differential 
geometry of the Bloch states and the topology of electromagnetic 
gauge fields in the form of a so-called axionic\cite{Wilczek1987} 
term, Eq.~\eqref{Eq:theta-term}, with $\theta$ representing  a uniform axion 
field. The axion is periodic and we find that for $\theta=\pi$ TRI holds, 
since under a time-reversal transformation $\theta\to -\theta$.\cite{QiHughesZhang}  

In terms of electric and magnetic field components, the axion term (\ref{Eq:theta-term}) 
becomes,
\begin{equation}
S_{\rm vpol }=\frac{e^2\theta}{4\pi^2}\int\!\text{d}^4x\,{\bf E}\cdot{\bf B}.
\end{equation} 
This magnetoelectric contribution is a topological term in real space, as it is more easily seen from the covariant 
writing, Eq.~\eqref{Eq:theta-term}, which clearly exhibits its independence of the metric. Furthermore, 
in view of Eq. (\ref{Eq:theta}) it is also topological in Bloch momentum space due to the  induced gauge structure 
in the Hilbert space of Bloch states. 
 
If TRI at the TI surface is broken in the presence of a magnetically ordered phase, then 
 ${\bf B}={\bf H}+4\pi{\bf M}$, and 
 a topological magnetoelectric effect (TME) has been predicted.\cite{QiHughesZhang, EssinMV} 
 This has inspired many proposals of 
 magnetic TI devices.\cite{GarateFranz, TserkovnyakLoss, SemenovDuanKim, YokoyamaZangNagaosa, NogueiraEremin, WangLianQiZhang, FerreirosBuijnstersKatsnelson, MorimotoFN2015} In the TME, an electric field causes a magnetic polarization in the same direction as the field. The TME is 
 the consequence of a CS term generated via the vacuum polarization due to proximity with a ferromagnetic insulator (FMI). If the FMI 
 is epitaxially grown on only one of the TI surfaces, there is only one CS term, in contrast to Eq. (\ref{Eq:diff-CS}).    
 The CS term yields an additional  Berry phase that modifies the dynamics of the magnetization.\cite{YokoyamaZangNagaosa, NogueiraEremin}

While previous studies have focused on the magnetic polarization generated by an externally applied electric field, in this paper we address a different important consequence of the TME, namely, its interplay with long-range Coulomb interaction among the Dirac electrons. The Coulomb interaction will generate a fluctuating electric field that interacts with the magnetization. 
Consequently, a non-local TME emerges that significantly impacts on the magnetization dynamics by an effective coupling over large distances.

Taking a similar approach as in Ref. \onlinecite{NogueiraEremin}, we will carry out explicit calculations of the vacuum polarization contributions to the effective action at zero temperature to leading order in the quantum fluctuations (one-loop diagrams) to derive the dynamics of both the magnetization and the Coulomb electric field at the TI/FMI interface.

\section{Model system}\label{SecModelSystem}
\begin{figure}
\includegraphics[width=0.5\columnwidth]{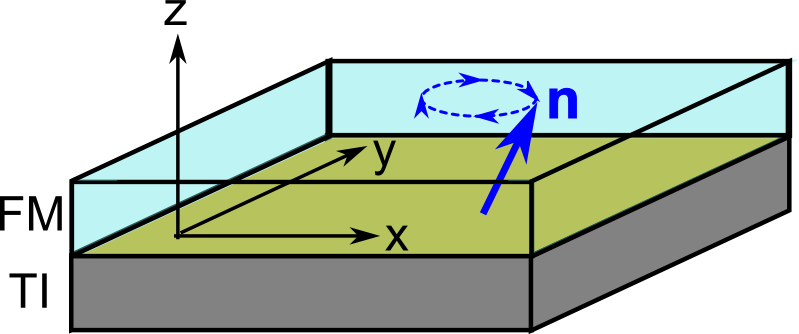}
\caption{Magnetization $\mathbf{n}$ at the interface of a topological insulator (TI) and a ferromagnetic insulator (FMI).}
\label{FigSketch}
\end{figure}

We consider the interface between a FMI layer on top of a TI, as shown in Fig.~\ref{FigSketch},
which we assume to lie in the $xy$ plane. As a starting point, we use on the one hand the 
Lagrangian density of a bulk FMI,\begin{equation}\label{EqFerromagnetLagrangian}
\mathcal{L}_\text{FMI} = \mathbf{b}\cdot\partial_t\mathbf{n} - \frac{\kappa}{2}\left[(\nablab\mathbf{n})^2+(\partial_z\mathbf{n})^2\right] -\frac{m^2}{2}\mathbf{n}^2 - \frac{u}{24}(\mathbf{n}^2)^2
,\end{equation}
where $\mathbf{n}$ is the magnetization, $\mathbf{b}$ is the Berry connection, $m^2<0$ for temperatures below the critical temperature of the magnetically ordered phase in the bulk, and $\kappa,u$ are positive constants.
Note that throughout this paper, $\nablab=(\partial_x, \partial_y, 0)$.

On the other hand, the topological Dirac electrons on the surface of a TI are described by
\begin{equation}\label{EqTopologicalInsulatorLagrangian}
\mathcal{L}_\text{TI} = \Psi^\dagger(\mathbf{r})\left[i\partial_t - iv_F(\sigma_y\partial_x-\sigma_x\partial_y) + J\boldsymbol{\sigma}\cdot\mathbf{n}(\mathbf{r})\right]\Psi(\mathbf{r})
,\end{equation}
where $\Psi^\dagger(\mathbf{r})$ creates an electron at position $\mathbf{r}$ in the $xy$ plane, $v_F$ is the Fermi velocity, $\boldsymbol{\sigma}=(\sigma_x,\sigma_y,\sigma_z)$ is the vector of Pauli matrices, and $J>0$ is the strength of the coupling of the electron spin to the magnetization $\mathbf{n}$ at $z=0$.

In addition, we account for long-range Coulomb interaction between the Dirac electrons at the interface,
\begin{equation}
V = \frac{1}{2}\sum_{\mathbf{q}}\rho(\mathbf{q})v_\text{Cou}(\mathbf{q})\rho(-\mathbf{q})
,\end{equation}
where the summation is over the two-dimensional momentum $\mathbf{q}$, the density operator is $\rho(\mathbf{q})=\sum_{\mathbf{k},s}\Psi^\dagger_{\mathbf{k}+\mathbf{q},s}\Psi_{\mathbf{k},s}$, with spin denoted $s$, and $v_\text{Cou}(\mathbf{q})$ is the Fourier transform of the Coulomb potential $v_\text{Cou}(\mathbf{r}-\mathbf{r}^\prime)=e^2/|\mathbf{r}-\mathbf{r}^\prime|$ for two electrons at positions $\mathbf{r}$ and $\mathbf{r}^\prime$, where $e$ is the elementary charge and the dielectric constant is $1/(4\pi)$ in Gaussian units. In two dimensions, the potential in reciprocal space takes 
the form
\begin{equation}
v_\text{Cou}(\mathbf{q}) = \frac{2\pi e^2}{|\mathbf{q}|}.
\end{equation}
The interaction can be made linear in electron density by a Hubbard-Stratonovich decoupling. With an auxiliary scalar field $\varphi$, that we define to have the unit of an electric potential, one finds the decoupled Lagrangian
\begin{equation}
\mathcal{L}_\text{Cou} = \sum_\mathbf{q}\left[e\varphi(\mathbf{q})\rho(\mathbf{q})-\frac{1}{4\pi}\varphi(-\mathbf{q})|\mathbf{q}|\varphi(\mathbf{q})\right]
,\end{equation}
which combined with Eq.~\eqref{EqTopologicalInsulatorLagrangian} gives the complete real-space Lagrangian density of conduction electrons at the interface,
\begin{eqnarray}
\mathcal{L}_\text{c}&=& \Psi^\dagger\!\left[i\partial_t + iv_F\hat{\mathbf{e}}_z\!\cdot\!(\boldsymbol{\sigma}\!\times\!\nablab) + J\boldsymbol{\sigma}\cdot\mathbf{n} +e\varphi\right]\!\Psi \notag\\
&&- \frac{1}{8\pi^2}[\nablab_\mathbf{r}\varphi(\mathbf{r})]\cdot\int\!\text{d}^2r^\prime\,\frac{\nablab_{\mathbf{r}^\prime}\varphi(\mathbf{r}^\prime)}{|\mathbf{r}-\mathbf{r}^\prime|}
.\end{eqnarray}
In total, the bilayer system is described by $\mathcal{L}=\mathcal{L}_\text{c}+\mathcal{L}_\text{FMI}$.


\section{Fluctuation effects}\label{SecFluctuations}
In this section, the quantum fluctuations will be evaluated to leading order by integrating out the electrons. First, we rewrite the fermionic part $\mathcal{L}^\text{f}_\text{c}$ of $\mathcal{L}_\text{c}$ in a form reminiscent of quantum electrodynamics\cite{NogueiraEremin}. With the definitions $\boldsymbol{\gamma}=(\gamma^0,\gamma^1,\gamma^2)=(\sigma_0,-i\sigma_x,-i\sigma_y)$, $\mathbf{a}=(\frac{e}{J}\varphi,n_y,-n_x)$, $\partial=(\partial_t,v_F\nablab)$, and the common notations $\overline{\Psi}=\Psi^\dagger\gamma^0$ and $\slashed{A}=\gamma^\mu A_\mu$, we get
\begin{equation}
\mathcal{L}_\text{c}^\text{f} = \overline{\Psi}\left[i\slashed{\partial}+J(n_z-\slashed{a})\right]\Psi
.\end{equation}
The mean-field value $\mathbf{n}_\text{MF}=\sigma_0\hat{\mathbf{e}}_z$ of the magnetization leads to an effective mass $m_{\Psi}=J\sigma_0$ of the fermion field, while $\tilde{\sigma}=n_z-\sigma_0$ describes the out-of-plane fluctuations. Integrating out the fermions in the standard way\cite{NegeleOrland} then leads to the action
\begin{equation}\label{EqFermionAction}
S_\text{c} = S_\text{MF} -\frac{J^2}{2}\text{Tr}[G(\tilde{\sigma}-\slashed{a})]^2
\end{equation}
with the propagator $G=(i\slashed{\partial}+m_{\Psi})^{-1}$. We relinquish an analysis of the mean-field action $S_\text{MF}$, which has been discussed in detail in Ref.~\onlinecite{NogueiraEremin}, and focus instead on the fluctuation effects. These are contained in the second term, $\delta S$, of Eq.~\eqref{EqFermionAction}, where we have already restricted ourselves to leading order. The operation $\text{Tr}$ implies integration over space-time and tracing out all quantum numbers. Diagrammatically, $\delta S$ contains four contributions to the vacuum polarization:
\unitlength = 1mm
\begin{fmffile}{diagrams}
  \fmfset{wiggly_len}{2mm}
	\fmfset{dash_len}{2mm}
  \fmfset{arrow_len}{2.5mm}
\begin{eqnarray}
\delta S &=\displaystyle\int\!\!\frac{\text{d}^3\lambda}{(2\pi)^3}&\left[
\raisebox{-0.22cm}{
\begin{fmfgraph}(22,6)
  \fmfleft{i}
	\fmfright{o}
	\fmf{wiggly, tension=0.6}{i,m1}
	\fmf{fermion, left, tension=0.4}{m1,m2,m1}
	\fmf{wiggly, tension=0.6}{m2,o}
\end{fmfgraph}} +
\raisebox{-0.22cm}{
\begin{fmfgraph}(22,6)
  \fmfleft{i}
	\fmfright{o}
	\fmf{dashes, tension=0.6}{i,m1}
	\fmf{fermion, left, tension=0.4}{m1,m2,m1}
	\fmf{dashes, tension=0.6}{m2,o}
\end{fmfgraph}} \right.\notag\\&&\left.+
\raisebox{-0.22cm}{
\begin{fmfgraph}(22,6)
  \fmfleft{i}
	\fmfright{o}
	\fmf{dashes, tension=0.6}{i,m1}
	\fmf{fermion, left, tension=0.4}{m1,m2,m1}
	\fmf{wiggly, tension=0.6}{m2,o}
\end{fmfgraph}} +
\raisebox{-0.22cm}{
\begin{fmfgraph}(22,6)
  \fmfleft{i}
	\fmfright{o}
	\fmf{wiggly, tension=0.6}{i,m1}
	\fmf{fermion, left, tension=0.4}{m1,m2,m1}
	\fmf{dashes, tension=0.6}{m2,o}
\end{fmfgraph}}\!\!\right]\!\!.\label{EqSgaugeDiagrammatic}
\end{eqnarray}
The fields $\Psi$, $a$, and $\tilde{\sigma}$ are represented by solid, wiggly, and dashed lines, respectively, and $\lambda$ comprises both frequency and momentum. Some details of the calculation of the diagrams can be found in Appendix~\ref{AppDiagrams}. Each of the mixed diagrams in the second line vanishes, and the remaining processes yield in the long-wavelength limit
\end{fmffile}
\begin{eqnarray}
\delta S &=& \frac{J^2}{8\pi}\int\!\text{d}t\int\!\text{d}^2r\notag\\&&
\left[(\mathbf{a}\times\partial)\cdot\mathbf{a} - \frac{(\partial\times\mathbf{a})^2}{3m_{\Psi}} -4m_\Psi\tilde{\sigma}^2 + \frac{(\partial\tilde{\sigma})^2}{3m_{\Psi}}\right]\label{EqDeltaAction}
\end{eqnarray}
Note that scalar products are to be taken in Minkowski space, with signature $(+,-,-)$.
As has been discussed earlier,\cite{NogueiraEremin, YokoyamaZangNagaosa} the term $(\mathbf{a}\times\partial)\cdot\mathbf{a}$ is a fluctuation-induced CS term. In total, we arrive at the following effective Lagrangian for the coupled FMI-TI bilayer system
\begin{eqnarray}
\mathcal{L}_\text{eff} &=& 
-\frac{\sigma_{xy}}{2 v_F^2}(\mathbf{n}\times\partial_t\mathbf{n})\cdot\hat{\mathbf{e}}_z
+\frac{\sigma_{xy}e}{v_F J}\mathbf{n}\cdot\nablab\varphi\notag\\
&&-\frac{NJ^2}{24\pi m_{\Psi}}\left[(\nablab\cdot\mathbf{n})^2+(\nablab n_z)^2\right]
+\frac{NJ^2}{24\pi v_F^2 m_{\Psi}}(\partial_t\mathbf{n})^2\notag\\
&&+\frac{Ne^2}{24\pi m_{\Psi}}(\nablab\varphi)^2
-\frac{NJe}{12\pi v_F m_{\Psi}}\left[(\nablab\varphi)\times(\partial_t\mathbf{n})\right]\cdot\hat{\mathbf{e}}_z\notag\\
&&-\frac{NJ^2m_{\Psi}}{2\pi v_F^2}n_z^2
+ \frac{NJm_{\Psi}^2}{\pi v_F^2}n_z\notag\\
&&+ \mathcal{L}_\text{FMI} - \frac{1}{8\pi^2}[\nablab_\mathbf{r}\varphi(\mathbf{r})]\cdot\int\!\text{d}^2r^\prime\,\frac{\nablab_{\mathbf{r}^\prime}\varphi(\mathbf{r}^\prime)}{|\mathbf{r}-\mathbf{r}^\prime|}\label{EqFullLagrangian}
,\end{eqnarray}
where $\hat{\mathbf{e}}_z$ is the unit vector in $z$ direction. Furthermore, we assumed $N$ orbital degrees of freedom of the Dirac electrons and defined the Hall conductance $\sigma_{xy}=NJ^2/(4\pi)$ in the two contributions from the CS term. The first one describes a Berry phase that adds up with the FMI Berry phase, while the second one leads to the TME. Derivatives of $\tilde{\sigma}$ have been replaced by derivatives of $n_z$, since $\sigma_0$ is constant.

Applying the Euler-Lagrange formalism on $\mathcal{L}_\text{eff}$ yields the Landau-Lifshitz equation (LLE) for the magnetization at the interface and the equation of motion for the fluctuating Coulomb potential $\varphi$. We arrange the LLE such that all first-order time derivatives of the magnetization are on the left side, such that it takes the form $A\cdot\partial_t\mathbf{n} = \mathbf{d}$ with a matrix $A$ and a vector $\mathbf{d}$ that depends on $\varphi$ and any other instance of $\mathbf{n}$. Since $A$ then collects precisely the Berry phase terms, it is antisymmetric and we can rewrite $A\cdot\partial_t\mathbf{n} = \mathbf{v}\times\partial_t\mathbf{n}$, where we find
\begin{equation}
\mathbf{v} = \frac{\mathbf{n}}{\mathbf{n}^2} + \frac{\sigma_{xy}}{v_F^2}\hat{\mathbf{e}}_z\label{EqBerryPhaseTerms}
.\end{equation}
The first term stems from the FMI Berry connection $\mathbf{b}$, which satisfies the condition $\partial_\mathbf{n}\times\mathbf{b}=-\mathbf{n}/\mathbf{n}^2$.
The second term originates with the CS term and enhances the overall Berry phase. If the magnetization is strong, the Berry phase may even be dominated by this topologically 
protected term. By taking the cross product with ${\bf n}$ in both sides of the 
equation ${\bf v}\times\partial_t{\bf n}={\bf d}$, we obtain, 
\begin{equation}
\label{Eq:LL-0}
\frac{{\bf v}}{2}\partial_t{\bf n}^2-({\bf n}\cdot{\bf v})\partial_t{\bf n}={\bf n}\times{\bf d}.
\end{equation}
Assuming that ${\bf n}^2$ is time-independent, Eq. \eqref{Eq:LL-0} becomes
\begin{equation}
\partial_t{\bf n}=\frac{{\bf d}\times{\bf n}}{1+\frac{\sigma_{xy}}{v_F^2}({\bf n}\cdot{\bf e}_z)}.
\label{EqLLEqResolved}
\end{equation}
We split $\mathbf{d}=\mathbf{d}_\mathbf{n}+\mathbf{d}_\varphi$ into the magnetization-dependent part
\begin{eqnarray}
\mathbf{d}_\mathbf{n} &=& \rho_\text{s}\cdot\nabla^2\mathbf{n}
+ \frac{NJ^2}{12\pi m_{\Psi}}\left[\frac{\partial_t^2\mathbf{n}}{v_F^2}+\nablab(\nablab\cdot\mathbf{n})\right]\notag\\
&&+\frac{NJm_{\Psi}}{\pi v_F^2}\left(Jn_z-m_{\Psi}\right)\hat{\mathbf{e}}_z
+\left(m^2+\frac{u}{6}\mathbf{n}^2\right)\!\mathbf{n}
,\end{eqnarray}
where the stiffness matrix is $\rho_\text{s}=\kappa 1\hspace{-0.25em}\mathrm{l}+(NJ^2/12\pi m_{\Psi})\text{diag}(0,0,1)$, and the contribution from the Coulomb interaction
\begin{eqnarray}
\mathbf{d}_\varphi &=&
-\frac{\sigma_{xy}e}{v_FJ}\nablab\varphi
-\frac{NJe}{12\pi v_Fm_{\Psi}}\hat{\mathbf{e}}_z\times\partial_t\nablab\varphi
\label{EqDphi}
.\end{eqnarray}
In addition, we obtain the Euler-Lagrange equation for the field $\varphi$. To make the physics more transparent, we write it in terms of the fluctuating electric field $\mathbf{E} = -\nablab\varphi$,
\begin{eqnarray}
0 &=& \frac{2\pi\sigma_{xy}e}{v_FJ}\mathbf{n}_\shortparallel + \frac{Ne}{6m_{\Psi}}\left(e\mathbf{E}-\frac{J}{v_F}\partial_t\mathbf{n}\times\hat{\mathbf{e}}_z\right)\notag\\
&& - \frac{1}{4\pi}\int\!\text{d}^2r^\prime\frac{\mathbf{E}(\mathbf{r}^\prime)}{|\mathbf{r}-\mathbf{r}^\prime|}\label{EqELEforEfield}
,\end{eqnarray}
where $\mathbf{n}_\shortparallel$ denotes the in-plane part of the magnetization.
This is an explicit form of the fluctuation-induced TME, where the electric field will be aligned with the magnetization, up to a dynamical correction depending on $\partial_t\mathbf{n}$. For the net field and magnetization this correction is irrelevant, since the time-average of $\partial_t\mathbf{n}$ vanishes. The first terms in Eqs.~\eqref{EqDphi} and \eqref{EqELEforEfield} stem from the contribution proportional to $\mathbf{n}\cdot\mathbf{E}$ in the Lagrangian, Eq.~\eqref{EqFullLagrangian}, representing the usual TME, which is a local effect. In contrast, the last term in Eq.~\eqref{EqELEforEfield} is a direct consequence of the long-range Coulomb interaction, and clearly makes the TME non-local by integration over the field at each point in the plane.

The motion of the magnetization becomes more clear when the bosonic field $\varphi$ in Eq.~\eqref{EqFullLagrangian} is integrated out as well. The part of the Lagrangian density that depends on the Coulomb interaction then becomes
\begin{equation}\label{EqLagrangianNoMorePhi}
\mathcal{L}_\varphi(\mathbf{r},t) = \frac{1}{2}\rho_{\bf n}(\mathbf{r},t)\int\!\text{d}^2r^\prime\frac{\rho_{\bf n}(\mathbf{r}^\prime,t)}{|\mathbf{r}-\mathbf{r}^\prime|},
\end{equation}
with the induced magnetic charge density,
\begin{equation}\label{EqTheMU}
\rho_{\bf n}(\mathbf{r},t) = \frac{\sigma_{xy}e}{v_FJ}\nablab\cdot\mathbf{n}(\mathbf{r},t)-\frac{NJe}{12\pi v_Fm_{\Psi}}\left[\nablab\times\partial_t\mathbf{n}(\mathbf{r},t)\right]\cdot\hat{\mathbf{e}}_z
.\end{equation}
Note that to leading order in momentum, the term involving $(\nabla\varphi)^2$ is negligible compared to the last term in Eq.~\eqref{EqFullLagrangian}. 
We observe that the fluctuation-induced magnetic charge contains an additional contribution besides the usual one. Typically, 
the magnetic charge density is proportional to $\nablab\cdot{\bf n}$ and usually arises in studies of magnetic skyrmions.\cite{NagaosaTokura}  
We also obtain a contribution $\sim (\nablab\times\partial_t{\bf n})\cdot\hat {\bf e}_z$, which does not have a topological origin. 
From the continuity equation we derive also the magnetic current density, 
\begin{eqnarray}
{\bf j}_{\bf n}&=&-\frac{\sigma_{xy}}{v_F}\partial_t{\bf n}
\nonumber\\
&&{}+\frac{NJe}{24\pi^2 v_Fm_{\Psi}}\int\! d^2r^\prime\frac{{\bf r}-{\bf r}^\prime}{|{\bf r}-{\bf r}^\prime|^2}
\left[\nablab_{\mathbf{r}^\prime}\!\times\partial_t\mathbf{n}(\mathbf{r}^\prime,t)\right]\cdot\hat{\mathbf{e}}_z.\nonumber\\
\end{eqnarray}

The magnetization dynamics is now determined by the integro-differential equation
\begin{equation}\label{EqLLEWithoutPhi}
\partial_t{\bf n}=\frac{{\bf D}_{\bf n}\times{\bf n}}{1+\frac{\sigma_{xy}}{v_F^2}({\bf n}\cdot{\bf e}_z)}
,\end{equation}
where,
\begin{equation}
\label{Eq:Dn}
{\bf D}_{\bf n}={\bf d}_{\bf n}+\frac{\sigma_{xy}e}{2v_FJ}\mathbf{E} + \frac{NJe}{24\pi v_Fm_{\Psi}}\hat{\mathbf{e}}_z\times\partial_t\mathbf{E},
\end{equation} 
and the electric field is now given explicitly by, 
\begin{equation}
\mathbf{E}(\mathbf{r}) =- \int\!\text{d}^2r^\prime\,\rho_{\bf n}(\mathbf{r}^\prime,t)\frac{(\mathbf{r}-\mathbf{r}^\prime)}{|\mathbf{r}-\mathbf{r}^\prime|^3}
.\end{equation}
The equation of motion can be simplified by an approximation of $\rho_{\bf n}$. 
Namely, in the low-frequency regime we can expect the second term in Eq.~\eqref{EqTheMU} to be small compared to the first term. 
Consequently, we find that the Coulomb interaction mainly acts via the CS term. The induced electric field 
 is then independent of $\partial_t\mathbf{n}$, and the equation can be brought into an explicit form similar to Eq.~\eqref{EqLLEqResolved}.

An important consequence of Eqs.~\eqref{EqTheMU} and \eqref{EqLLEWithoutPhi} is that the Coulomb interaction does not directly couple the magnetizations at different points in the plane. Rather, 
it is the {\it divergence} of the magnetization that enters into the magnetization dynamics over long distances. This can be understood by the duality of magnetic and electric charges on the 
surface of a TI,\cite{NomuraNagaosa} where $\nabla\cdot\mathbf{n}$ is equivalent to an electric charge of the magnetic texture. This charge generates a Coulomb field.
In the case of a uniform magnetization, where both $\rho_{\bf n}$ and $\mathbf{E}$ are absent, the Coulomb interaction will thus not affect the magnetization dynamics. We are then left with the LLE \eqref{EqLLEqResolved} with ${\bf d}={\bf d}_{\bf n}$, 
where also in $\mathbf{d}_\mathbf{n}$, all spatial derivatives vanish. From the remaining terms, we simply obtain a precession of the magnetization around the $z$ axis by Eq.~\eqref{EqLLEqResolved}.

To illustrate how the long-range Coulomb interaction affects the dynamics, we turn to a simple example of a nonuniform magnetization. Assume that the system is prepared with a magnetic texture, where the phase of the precession changes within a narrow region about $x=0$. The divergence of $\mathbf{n}$ will then be nonzero within that region. The corresponding terms in $\mathbf{d}_\mathbf{n}$ will locally alter the magnetization dynamics, trying to align the magnetization at neighboring sites. This will smoothen the transition at $x=0$ and evoke spin waves spreading in both half-planes. However, via the Coulomb interaction, there is an instantaneous impact on the magnetization even far from the texture.
For large $x$, we can assume $\rho_{\bf n}=\rho_{{\bf n},0}\delta(x)$, where 
$\rho_{{\bf n},0}$ oscillates with the precession frequency at $x=0$, and one readily verifies that 
$\mathbf{E}=2\hat{\mathbf{e}}_x\rho_{{\bf n},0}/x$. The second and third term in Eq.~\eqref{Eq:Dn} lead to in-plane components of the effective field of precession in $x$ and $y$ direction, respectively, where the latter can be neglected in the low-frequency limit. Thus, the effective field at arbitrary $x$ is already tilted away from the $z$ direction before the spin waves due to the local stiffness terms arrive.

As a final remark, we note that only the in-plane inhomogeneities of the magnetization participate in the Coulomb driven dynamics, while the out-of-plane magnetization does not enter. Therefore, we can expect a similar non-local effect if we replace the texture discussed above by a domain wall, 
as long as the rotation of the magnetization within the transition region happens in a way that involves an in-plane divergence. Apart from evoking Coulomb terms in the magnetization dynamics, 
the presence of a domain wall in a magnetic layer on a TI also leads to other effects, e.g., chiral currents, that we have not discussed in this paper, but have been subject to a number of previous studies.\cite{TserkovnyakLoss, Linder2014, FerreirosBuijnstersKatsnelson, FerreirosCortijo, WicklesBelzig}

\section{Conclusion}
We have analytically studied a TI-FMI interface in the presence of long-range Coulomb interaction and derived the fluctuation-induced dynamics of both the magnetization and the electric field mediating Coulomb interactions, to second order in gradients and fields. We have found that, as 
a result of long-range interactions, the TME becomes non-local, such that the magnetization is
 coupled to the electric field anywhere in the plane. The CS term in the effective action enhances the overall Berry phase and thus modifies the magnitude of the effective field of the magnetization precession. Magnetic textures involving a divergence of the in-plane magnetization tilt the effective field of the precession in a non-local way.

\begin{acknowledgments}
A.S. and S.R. acknowledge support from the Norwegian Research Council, 
Grants 205591/V20 and 216700/F20. F.S.N. acknowledges support from the Collaborative 
Research Center SFB 1143 "Correlated Magnetism: From Frustration to Topology".
\end{acknowledgments}

\appendix

\section{Calculation of the diagrams}\label{AppDiagrams}
In this appendix, we present the zero-temperature calculation that leads to Eq.~\eqref{EqDeltaAction}. From Eq.~\eqref{EqFermionAction}, we have
\begin{equation}
\delta S = -\frac{J^2}{2}\int\!\text{d}t\int\!\text{d}^2x\,\sum_{\varkappa}\langle\varkappa|\text{tr}[G(\tilde{\sigma}-\slashed{a})]^2|\varkappa\rangle
,\end{equation}
where the trace $\text{tr}$ is taken in spin space, $\varkappa$ denotes all other quantum numbers, and the propagator is
\begin{equation}
G = \frac{-i\slashed{\partial}+m_\Psi}{\partial^2+m_\Psi^2}
.\end{equation}
We go to imaginary time by Wick rotation, $\tau=it$, which makes space-time Euclidean. The Dirac $\gamma$-matrices are then identical to the Pauli matrices. We find $i\slashed{\partial}\rightarrow-\slashed{\partial}$ and $\slashed{a}\rightarrow\slashed{\alpha}$, where we defined $\alpha=(a^0,ia^1,ia^2)$. Furthermore, $\delta S$ is transformed to reciprocal space and the sum over electron quantum numbers is carried out in a basis of plane-wave states, $\varkappa=(\omega,\mathbf{k})$, with frequency $\omega$ and twodimensional momentum $\mathbf{k}$. The frequency and momentum of the bosonic fields $\alpha$ and $\tilde{\sigma}$ in reciprocal space, are denoted $\lambda=(\Omega, \mathbf{q})$. We get
\begin{widetext}
\begin{equation}
\delta S = \frac{iJ^2}{2}\int\!\frac{\text{d}^3\lambda}{(2\pi)^3}\int\!\frac{\text{d}^3\varkappa}{(2\pi)^3}\,\frac{\text{tr}\left[(m_\Psi+i\slashed{\varkappa})(-\slashed{\alpha}(\lambda)+\tilde{\sigma}(\lambda))(m_\Psi+i(\slashed{\varkappa}-\slashed{\lambda}))(-\slashed{\alpha}(-\lambda)+\tilde{\sigma}(-\lambda))\right]}{(\varkappa^2+m_\Psi^2)((\varkappa-\lambda)^2+m_\Psi^2)}\label{EqDSgaugeNotYetMultipliedOut}
,\end{equation}
\end{widetext}
and the matrix structure inside the remaining trace is now determined by products of Dirac matrices. As one can easily verify by using the commutation  and anticommutation relations of the Euclidean Dirac matrices\cite{ItzZuber, NogueiraEremin2013}, $\text{tr}(\gamma_\mu\gamma_\nu)=2\delta_{\mu\nu}$, $\text{tr}(\gamma_\mu\gamma_\nu\gamma_\lambda)=2i\varepsilon_{\mu\nu\lambda}$, and $\text{tr}(\gamma_\mu\gamma_\nu\gamma_\lambda\gamma_\rho)=2(\delta_{\mu\nu}\delta_{\lambda\rho}-\delta_{\mu\lambda}\delta_{\nu\rho}+\delta_{\mu\rho}\delta_{\nu\lambda})$. Inserting these formulas into the numerator of the integrand in Eq.~\eqref{EqDSgaugeNotYetMultipliedOut} yields
\begin{eqnarray}
\lefteqn{\text{tr}[\hdots] =}\notag\\
&&2\alpha_\mu(\lambda)\alpha_\nu(-\lambda)\big[m_\Psi\varepsilon_{\mu\rho\nu}\lambda_\rho+ \delta_{\mu\nu}(m_\Psi^2+\varkappa\cdot(\varkappa-\lambda))\notag\\&&\hspace{2.4cm}-2\varkappa_\mu\varkappa_\nu + \varkappa_\nu\lambda_\mu + \varkappa_\mu\lambda_\nu\big] \notag\\
&& +2\tilde{\sigma}(\lambda)\tilde{\sigma}(-\lambda)\big[m_\Psi^2-\varkappa\cdot(\varkappa-\lambda)\big]\notag\\
&& +2i\tilde{\sigma}(\lambda)\alpha_\mu(-\lambda)\big[-m_\Psi(\varkappa_\mu-\lambda_\mu)-m_\Psi\varkappa_\mu\notag\\&&\hspace{2.7cm}+\varepsilon_{\rho\nu\mu}\varkappa_\rho(\varkappa_\nu-\lambda_\nu)\big]\notag\\
&& +2i\alpha_\mu(\lambda)\tilde{\sigma}(-\lambda)\big[-m_\Psi(\varkappa_\mu-\lambda_\mu)-m_\Psi\varkappa_\mu\notag\\&&\hspace{2.7cm}+\varepsilon_{\rho\mu\nu}\varkappa_\rho(\varkappa_\nu-\lambda_\nu)\big]\label{EqDSgaugeMultipliedOut}
,\end{eqnarray}
corresponding to the four diagrams in Eq.~\eqref{EqSgaugeDiagrammatic}. Let these diagrams be called $D_1,\hdots,D_4$, in the same order as in Eq.~\eqref{EqSgaugeDiagrammatic}.
Next, the integral over $\varkappa$ will be carried out. As has been disussed in Appendix~A of Ref.~\cite{NogueiraEremin2013}, one can rewrite the first diagram to take the form
\begin{eqnarray}
D_1(\lambda)
&=& iJ^2a_\mu(\lambda)a_\nu(-\lambda)\Bigg[\varepsilon_{\mu\rho\nu}m_{\Psi}\lambda_\rho I_1(\lambda)\notag\\&&+ P_{\mu\nu}(\lambda)\!\left(m_{\Psi}I_1(\lambda)-\frac{\lambda^2}{4}I_1(\lambda)+\frac{1}{2}I_2\right)\!\!\Bigg]
,\end{eqnarray}
with the projector $P_{\mu\nu}(\lambda)=\delta_{\mu\nu}-\lambda_\mu\lambda_\nu/\lambda^2$ and the integrals
\begin{eqnarray}
I_1(\lambda) &=& \int\!\!\frac{d^3\varkappa}{(2\pi)^3}\,\frac{1}{(\varkappa^2+m_\Psi^2)((\varkappa-\lambda)^2+m_\Psi^2)}\notag\\
&=& \frac{1}{4\pi|\lambda|}\arctan\left(\frac{|\lambda|}{2m_\Psi}\right)
,\end{eqnarray}
\begin{equation}
I_2 = \int\frac{\text{d}^3\varkappa}{(2\pi)^3}\,\frac{1}{\varkappa^2+m_\Psi^2}
 = -\frac{m_\Psi}{4\pi}
,\end{equation}
with the result
\begin{equation}
D_1(\lambda) = iNJ^2\alpha_\mu(\lambda)\alpha_\nu(-\lambda)\left[\frac{\varepsilon_{\mu\rho\nu}\lambda_\rho}{8\pi}-\frac{\lambda^2P_{\mu\nu}(\lambda)}{24\pi m_\Psi}\right]
\end{equation}
to second order in $\lambda$. Note that $I_2$ requires dimensional regularization\cite{ItzZuber}, since it is formally divergent. By simple manipulations, one can reduce the second diagram to the same integrals:
\begin{eqnarray}
D_2(\lambda) &=& iNJ^2\tilde{\sigma}(\lambda)\tilde{\sigma}(-\lambda)\left[\left(2m_\Psi^2+\frac{1}{2}\lambda^2\right)I_1(\lambda) - I_2\right]\notag\\
&=& iNJ^2\tilde{\sigma}(\lambda)\tilde{\sigma}(-\lambda)\left[\frac{m_\Psi}{2\pi} + \frac{\lambda^2}{24\pi m_\Psi}\right] + \mathcal{O}\left(\lambda^3\right)
\end{eqnarray}
In the two diagrams mixing $\alpha$ and $\tilde{\sigma}$, performing the $\varkappa$ integration leads to $D_3(\lambda)=D_4(\lambda)=0$. Summing up the contributions from $D_1$ and $D_2$ and transforming back to real space and real time finally yields Eq.~\eqref{EqDeltaAction}.

\bibliography{references}

\providecommand{\noopsort}[1]{}\providecommand{\singleletter}[1]{#1}%
\begin{thebibliography}{25}%
\makeatletter
\providecommand \@ifxundefined [1]{%
 \@ifx{#1\undefined}
}%
\providecommand \@ifnum [1]{%
 \ifnum #1\expandafter \@firstoftwo
 \else \expandafter \@secondoftwo
 \fi
}%
\providecommand \@ifx [1]{%
 \ifx #1\expandafter \@firstoftwo
 \else \expandafter \@secondoftwo
 \fi
}%
\providecommand \natexlab [1]{#1}%
\providecommand \enquote  [1]{``#1''}%
\providecommand \bibnamefont  [1]{#1}%
\providecommand \bibfnamefont [1]{#1}%
\providecommand \citenamefont [1]{#1}%
\providecommand \href@noop [0]{\@secondoftwo}%
\providecommand \href [0]{\begingroup \@sanitize@url \@href}%
\providecommand \@href[1]{\@@startlink{#1}\@@href}%
\providecommand \@@href[1]{\endgroup#1\@@endlink}%
\providecommand \@sanitize@url [0]{\catcode `\\12\catcode `\$12\catcode
  `\&12\catcode `\#12\catcode `\^12\catcode `\_12\catcode `\%12\relax}%
\providecommand \@@startlink[1]{}%
\providecommand \@@endlink[0]{}%
\providecommand \url  [0]{\begingroup\@sanitize@url \@url }%
\providecommand \@url [1]{\endgroup\@href {#1}{\urlprefix }}%
\providecommand \urlprefix  [0]{URL }%
\providecommand \Eprint [0]{\href }%
\providecommand \doibase [0]{http://dx.doi.org/}%
\providecommand \selectlanguage [0]{\@gobble}%
\providecommand \bibinfo  [0]{\@secondoftwo}%
\providecommand \bibfield  [0]{\@secondoftwo}%
\providecommand \translation [1]{[#1]}%
\providecommand \BibitemOpen [0]{}%
\providecommand \bibitemStop [0]{}%
\providecommand \bibitemNoStop [0]{.\EOS\space}%
\providecommand \EOS [0]{\spacefactor3000\relax}%
\providecommand \BibitemShut  [1]{\csname bibitem#1\endcsname}%
\let\auto@bib@innerbib\@empty
\bibitem [{\citenamefont {Hasan}\ and\ \citenamefont {Kane}(2010)}]{Kane-RMP}%
  \BibitemOpen
  \bibfield  {author} {\bibinfo {author} {\bibfnamefont {M.~Z.}\ \bibnamefont
  {Hasan}}\ and\ \bibinfo {author} {\bibfnamefont {C.}~\bibnamefont {Kane}},\
  }\href@noop {} {\bibfield  {journal} {\bibinfo  {journal} {Rev.\ Mod.\ Phys}\
  }\textbf {\bibinfo {volume} {82}},\ \bibinfo {pages} {3045} (\bibinfo {year}
  {2010})}\BibitemShut {NoStop}%
\bibitem [{\citenamefont {Qi}\ and\ \citenamefont {Zhang}(2011)}]{Zhang-RMP}%
  \BibitemOpen
  \bibfield  {author} {\bibinfo {author} {\bibfnamefont {X.-L.}\ \bibnamefont
  {Qi}}\ and\ \bibinfo {author} {\bibfnamefont {S.-C.}\ \bibnamefont {Zhang}},\
  }\href@noop {} {\bibfield  {journal} {\bibinfo  {journal} {Rev.\ Mod.\ Phys}\
  }\textbf {\bibinfo {volume} {83}},\ \bibinfo {pages} {1057} (\bibinfo {year}
  {2011})}\BibitemShut {NoStop}%
\bibitem [{\citenamefont {Qi}\ \emph {et~al.}(2008{\natexlab{a}})\citenamefont
  {Qi}, \citenamefont {Hughes},\ and\ \citenamefont
  {Zhang}}]{QiHughesZhang_Nature}%
  \BibitemOpen
  \bibfield  {author} {\bibinfo {author} {\bibfnamefont {X.-L.}\ \bibnamefont
  {Qi}}, \bibinfo {author} {\bibfnamefont {T.~L.}\ \bibnamefont {Hughes}}, \
  and\ \bibinfo {author} {\bibfnamefont {S.-C.}\ \bibnamefont {Zhang}},\
  }\href@noop {} {\bibfield  {journal} {\bibinfo  {journal} {Nature\ Phys.}\
  }\textbf {\bibinfo {volume} {4}},\ \bibinfo {pages} {273} (\bibinfo {year}
  {2008}{\natexlab{a}})}\BibitemShut {NoStop}%
\bibitem [{\citenamefont {Yokoyama}\ \emph
  {et~al.}(2010{\natexlab{a}})\citenamefont {Yokoyama}, \citenamefont
  {Tanaka},\ and\ \citenamefont {Nagaosa}}]{YokoyamaTanakaNagaosa}%
  \BibitemOpen
  \bibfield  {author} {\bibinfo {author} {\bibfnamefont {T.}~\bibnamefont
  {Yokoyama}}, \bibinfo {author} {\bibfnamefont {Y.}~\bibnamefont {Tanaka}}, \
  and\ \bibinfo {author} {\bibfnamefont {N.}~\bibnamefont {Nagaosa}},\
  }\href@noop {} {\bibfield  {journal} {\bibinfo  {journal} {Phys.\ Rev.\ B}\
  }\textbf {\bibinfo {volume} {81}},\ \bibinfo {pages} {121401(R)} (\bibinfo
  {year} {2010}{\natexlab{a}})}\BibitemShut {NoStop}%
\bibitem [{\citenamefont {Qi}\ \emph {et~al.}(2008{\natexlab{b}})\citenamefont
  {Qi}, \citenamefont {Hughes},\ and\ \citenamefont {Zhang}}]{QiHughesZhang}%
  \BibitemOpen
  \bibfield  {author} {\bibinfo {author} {\bibfnamefont {X.-L.}\ \bibnamefont
  {Qi}}, \bibinfo {author} {\bibfnamefont {T.~L.}\ \bibnamefont {Hughes}}, \
  and\ \bibinfo {author} {\bibfnamefont {S.-C.}\ \bibnamefont {Zhang}},\
  }\href@noop {} {\bibfield  {journal} {\bibinfo  {journal} {Phys.\ Rev.\ B}\
  }\textbf {\bibinfo {volume} {78}},\ \bibinfo {pages} {195424} (\bibinfo
  {year} {2008}{\natexlab{b}})}\BibitemShut {NoStop}%
\bibitem [{\citenamefont {Essin}\ \emph {et~al.}(2009)\citenamefont {Essin},
  \citenamefont {Moore},\ and\ \citenamefont {Vanderbilt}}]{EssinMV}%
  \BibitemOpen
  \bibfield  {author} {\bibinfo {author} {\bibfnamefont {A.~M.}\ \bibnamefont
  {Essin}}, \bibinfo {author} {\bibfnamefont {J.~E.}\ \bibnamefont {Moore}}, \
  and\ \bibinfo {author} {\bibfnamefont {D.}~\bibnamefont {Vanderbilt}},\
  }\href@noop {} {\bibfield  {journal} {\bibinfo  {journal} {Phys.\ Rev.\
  Lett.}\ }\textbf {\bibinfo {volume} {102}},\ \bibinfo {pages} {146805}
  (\bibinfo {year} {2009})}\BibitemShut {NoStop}%
\bibitem [{\citenamefont {Niemi}\ and\ \citenamefont
  {Semenoff}(1983)}]{Semenoff}%
  \BibitemOpen
  \bibfield  {author} {\bibinfo {author} {\bibfnamefont {A.~J.}\ \bibnamefont
  {Niemi}}\ and\ \bibinfo {author} {\bibfnamefont {G.~W.}\ \bibnamefont
  {Semenoff}},\ }\href@noop {} {\bibfield  {journal} {\bibinfo  {journal}
  {Phys.\ Rev.\ Lett.}\ }\textbf {\bibinfo {volume} {51}},\ \bibinfo {pages}
  {2077} (\bibinfo {year} {1983})}\BibitemShut {NoStop}%
\bibitem [{\citenamefont {Redlich}(1984)}]{Redlich}%
  \BibitemOpen
  \bibfield  {author} {\bibinfo {author} {\bibfnamefont {A.~N.}\ \bibnamefont
  {Redlich}},\ }\href@noop {} {\bibfield  {journal} {\bibinfo  {journal}
  {Phys.\ Rev. D}\ }\textbf {\bibinfo {volume} {29}},\ \bibinfo {pages} {2366}
  (\bibinfo {year} {1984})}\BibitemShut {NoStop}%
\bibitem [{\citenamefont {Wilczek}(1987)}]{Wilczek1987}%
  \BibitemOpen
  \bibfield  {author} {\bibinfo {author} {\bibfnamefont {F.}~\bibnamefont
  {Wilczek}},\ }\href@noop {} {\bibfield  {journal} {\bibinfo  {journal}
  {Phys.\ Rev.\ Lett.}\ }\textbf {\bibinfo {volume} {58}},\ \bibinfo {pages}
  {1799} (\bibinfo {year} {1987})}\BibitemShut {NoStop}%
\bibitem [{\citenamefont {Garate}\ and\ \citenamefont
  {Franz}(2010)}]{GarateFranz}%
  \BibitemOpen
  \bibfield  {author} {\bibinfo {author} {\bibfnamefont {I.}~\bibnamefont
  {Garate}}\ and\ \bibinfo {author} {\bibfnamefont {M.}~\bibnamefont {Franz}},\
  }\href@noop {} {\bibfield  {journal} {\bibinfo  {journal} {Phys.\ Rev.\
  Lett.}\ }\textbf {\bibinfo {volume} {104}},\ \bibinfo {pages} {146802}
  (\bibinfo {year} {2010})}\BibitemShut {NoStop}%
\bibitem [{\citenamefont {Tserkovnyak}\ and\ \citenamefont
  {Loss}(2012)}]{TserkovnyakLoss}%
  \BibitemOpen
  \bibfield  {author} {\bibinfo {author} {\bibfnamefont {Y.}~\bibnamefont
  {Tserkovnyak}}\ and\ \bibinfo {author} {\bibfnamefont {D.}~\bibnamefont
  {Loss}},\ }\href@noop {} {\bibfield  {journal} {\bibinfo  {journal} {Phys.\
  Rev.\ Lett.}\ }\textbf {\bibinfo {volume} {108}},\ \bibinfo {pages} {187201}
  (\bibinfo {year} {2012})}\BibitemShut {NoStop}%
\bibitem [{\citenamefont {Semenov}\ \emph {et~al.}(2012)\citenamefont
  {Semenov}, \citenamefont {Duan},\ and\ \citenamefont {Kim}}]{SemenovDuanKim}%
  \BibitemOpen
  \bibfield  {author} {\bibinfo {author} {\bibfnamefont {Y.~G.}\ \bibnamefont
  {Semenov}}, \bibinfo {author} {\bibfnamefont {X.}~\bibnamefont {Duan}}, \
  and\ \bibinfo {author} {\bibfnamefont {K.~W.}\ \bibnamefont {Kim}},\
  }\href@noop {} {\bibfield  {journal} {\bibinfo  {journal} {Phys.\ Rev.\ B}\
  }\textbf {\bibinfo {volume} {86}},\ \bibinfo {pages} {161406(R)} (\bibinfo
  {year} {2012})}\BibitemShut {NoStop}%
\bibitem [{\citenamefont {Yokoyama}\ \emph
  {et~al.}(2010{\natexlab{b}})\citenamefont {Yokoyama}, \citenamefont {Zang},\
  and\ \citenamefont {Nagaosa}}]{YokoyamaZangNagaosa}%
  \BibitemOpen
  \bibfield  {author} {\bibinfo {author} {\bibfnamefont {T.}~\bibnamefont
  {Yokoyama}}, \bibinfo {author} {\bibfnamefont {J.}~\bibnamefont {Zang}}, \
  and\ \bibinfo {author} {\bibfnamefont {N.}~\bibnamefont {Nagaosa}},\
  }\href@noop {} {\bibfield  {journal} {\bibinfo  {journal} {Phys.\ Rev.\ B}\
  }\textbf {\bibinfo {volume} {81}},\ \bibinfo {pages} {241410(R)} (\bibinfo
  {year} {2010}{\natexlab{b}})}\BibitemShut {NoStop}%
\bibitem [{\citenamefont {Nogueira}\ and\ \citenamefont
  {Eremin}(2012)}]{NogueiraEremin}%
  \BibitemOpen
  \bibfield  {author} {\bibinfo {author} {\bibfnamefont {F.~S.}\ \bibnamefont
  {Nogueira}}\ and\ \bibinfo {author} {\bibfnamefont {I.}~\bibnamefont
  {Eremin}},\ }\href@noop {} {\bibfield  {journal} {\bibinfo  {journal} {Phys.\
  Rev.\ Lett.}\ }\textbf {\bibinfo {volume} {109}},\ \bibinfo {pages} {237203}
  (\bibinfo {year} {2012})}\BibitemShut {NoStop}%
\bibitem [{\citenamefont {Wang}\ \emph {et~al.}(2015)\citenamefont {Wang},
  \citenamefont {Lian}, \citenamefont {Qi},\ and\ \citenamefont
  {Zhang}}]{WangLianQiZhang}%
  \BibitemOpen
  \bibfield  {author} {\bibinfo {author} {\bibfnamefont {J.}~\bibnamefont
  {Wang}}, \bibinfo {author} {\bibfnamefont {B.}~\bibnamefont {Lian}}, \bibinfo
  {author} {\bibfnamefont {X.-L.}\ \bibnamefont {Qi}}, \ and\ \bibinfo {author}
  {\bibfnamefont {S.-C.}\ \bibnamefont {Zhang}},\ }\href@noop {} {\bibfield
  {journal} {\bibinfo  {journal} {Phys.\ Rev.\ B}\ }\textbf {\bibinfo {volume}
  {92}},\ \bibinfo {pages} {081107(R)} (\bibinfo {year} {2015})}\BibitemShut
  {NoStop}%
\bibitem [{\citenamefont {Ferreiros}\ \emph {et~al.}(2015)\citenamefont
  {Ferreiros}, \citenamefont {Buijnsters},\ and\ \citenamefont
  {Katsnelson}}]{FerreirosBuijnstersKatsnelson}%
  \BibitemOpen
  \bibfield  {author} {\bibinfo {author} {\bibfnamefont {Y.}~\bibnamefont
  {Ferreiros}}, \bibinfo {author} {\bibfnamefont {F.~J.}\ \bibnamefont
  {Buijnsters}}, \ and\ \bibinfo {author} {\bibfnamefont {M.~I.}\ \bibnamefont
  {Katsnelson}},\ }\href@noop {} {\bibfield  {journal} {\bibinfo  {journal}
  {Phys.\ Rev.\ B}\ }\textbf {\bibinfo {volume} {92}},\ \bibinfo {pages}
  {085416} (\bibinfo {year} {2015})}\BibitemShut {NoStop}%
\bibitem [{\citenamefont {Morimoto}\ \emph {et~al.}(2015)\citenamefont
  {Morimoto}, \citenamefont {Furusaki},\ and\ \citenamefont
  {Nagaosa}}]{MorimotoFN2015}%
  \BibitemOpen
  \bibfield  {author} {\bibinfo {author} {\bibfnamefont {T.}~\bibnamefont
  {Morimoto}}, \bibinfo {author} {\bibfnamefont {A.}~\bibnamefont {Furusaki}},
  \ and\ \bibinfo {author} {\bibfnamefont {N.}~\bibnamefont {Nagaosa}},\
  }\href@noop {} {\bibfield  {journal} {\bibinfo  {journal} {Phys.\ Rev.\ B}\
  }\textbf {\bibinfo {volume} {92}},\ \bibinfo {pages} {085113} (\bibinfo
  {year} {2015})}\BibitemShut {NoStop}%
\bibitem [{\citenamefont {Negele}\ and\ \citenamefont
  {Orland}(1988)}]{NegeleOrland}%
  \BibitemOpen
  \bibfield  {author} {\bibinfo {author} {\bibfnamefont {J.~W.}\ \bibnamefont
  {Negele}}\ and\ \bibinfo {author} {\bibfnamefont {H.}~\bibnamefont
  {Orland}},\ }\href@noop {} {\emph {\bibinfo {title} {Quantum Many-particle
  systems}}}\ (\bibinfo  {publisher} {Addison-Wesley},\ \bibinfo {year}
  {1988})\BibitemShut {NoStop}%
\bibitem [{\citenamefont {Nagaosa}\ and\ \citenamefont
  {Tokura}(2013)}]{NagaosaTokura}%
  \BibitemOpen
  \bibfield  {author} {\bibinfo {author} {\bibfnamefont {N.}~\bibnamefont
  {Nagaosa}}\ and\ \bibinfo {author} {\bibfnamefont {Y.}~\bibnamefont
  {Tokura}},\ }\href@noop {} {\bibfield  {journal} {\bibinfo  {journal}
  {Nature\ Nanotech.}\ }\textbf {\bibinfo {volume} {8}},\ \bibinfo {pages}
  {899} (\bibinfo {year} {2013})}\BibitemShut {NoStop}%
\bibitem [{\citenamefont {Nomura}\ and\ \citenamefont
  {Nagaosa}(2010)}]{NomuraNagaosa}%
  \BibitemOpen
  \bibfield  {author} {\bibinfo {author} {\bibfnamefont {K.}~\bibnamefont
  {Nomura}}\ and\ \bibinfo {author} {\bibfnamefont {N.}~\bibnamefont
  {Nagaosa}},\ }\href@noop {} {\bibfield  {journal} {\bibinfo  {journal}
  {Phys.\ Rev.\ B}\ }\textbf {\bibinfo {volume} {82}},\ \bibinfo {pages}
  {161401(R)} (\bibinfo {year} {2010})}\BibitemShut {NoStop}%
\bibitem [{\citenamefont {Linder}(2014)}]{Linder2014}%
  \BibitemOpen
  \bibfield  {author} {\bibinfo {author} {\bibfnamefont {J.}~\bibnamefont
  {Linder}},\ }\href@noop {} {\bibfield  {journal} {\bibinfo  {journal} {Phys.\
  Rev.\ B}\ }\textbf {\bibinfo {volume} {90}},\ \bibinfo {pages} {041412(R)}
  (\bibinfo {year} {2014})}\BibitemShut {NoStop}%
\bibitem [{\citenamefont {Ferreiros}\ and\ \citenamefont
  {Cortijo}(2014)}]{FerreirosCortijo}%
  \BibitemOpen
  \bibfield  {author} {\bibinfo {author} {\bibfnamefont {Y.}~\bibnamefont
  {Ferreiros}}\ and\ \bibinfo {author} {\bibfnamefont {A.}~\bibnamefont
  {Cortijo}},\ }\href@noop {} {\bibfield  {journal} {\bibinfo  {journal}
  {Phys.\ Rev.\ B}\ }\textbf {\bibinfo {volume} {89}},\ \bibinfo {pages}
  {024413} (\bibinfo {year} {2014})}\BibitemShut {NoStop}%
\bibitem [{\citenamefont {Wickles}\ and\ \citenamefont
  {Belzig}(2012)}]{WicklesBelzig}%
  \BibitemOpen
  \bibfield  {author} {\bibinfo {author} {\bibfnamefont {C.}~\bibnamefont
  {Wickles}}\ and\ \bibinfo {author} {\bibfnamefont {W.}~\bibnamefont
  {Belzig}},\ }\href@noop {} {\bibfield  {journal} {\bibinfo  {journal} {Phys.\
  Rev.\ B}\ }\textbf {\bibinfo {volume} {86}},\ \bibinfo {pages} {035151}
  (\bibinfo {year} {2012})}\BibitemShut {NoStop}%
\bibitem [{\citenamefont {Itzykson}\ and\ \citenamefont
  {Zuber}(1980)}]{ItzZuber}%
  \BibitemOpen
  \bibfield  {author} {\bibinfo {author} {\bibfnamefont {C.}~\bibnamefont
  {Itzykson}}\ and\ \bibinfo {author} {\bibfnamefont {J.-B.}\ \bibnamefont
  {Zuber}},\ }\href@noop {} {\emph {\bibinfo {title} {Quantum Field Theory}}}\
  (\bibinfo  {publisher} {Mcgraw-Hill College},\ \bibinfo {year}
  {1980})\BibitemShut {NoStop}%
\bibitem [{\citenamefont {Nogueira}\ and\ \citenamefont
  {Eremin}(2013)}]{NogueiraEremin2013}%
  \BibitemOpen
  \bibfield  {author} {\bibinfo {author} {\bibfnamefont {F.~S.}\ \bibnamefont
  {Nogueira}}\ and\ \bibinfo {author} {\bibfnamefont {I.}~\bibnamefont
  {Eremin}},\ }\href@noop {} {\bibfield  {journal} {\bibinfo  {journal} {Phys.\
  Rev. B}\ }\textbf {\bibinfo {volume} {88}},\ \bibinfo {pages} {085126}
  (\bibinfo {year} {2013})}\BibitemShut {NoStop}%
\end{thebibliography}%

\end{document}